\documentclass[]{ws-procs9x6}            
\usepackage{slashed}

\begin{document}

\title{$\Delta(1232)$ Contribution in the $\pi$-photoproduction on Nucleons \\
in Covariant Chiral Perturbation Theory}

\author{G. H. Guerrero Navarro$^*$ 
}

\address{Departamento de F\'{\i}sica Teorica and Instituto de Fisica Corpuscular (IFIC),
Centro Mixto UVEG-CSIC,\\
Valencia E-46071, Spain\\
$^*$E-mail: gusguena@ific.uv.es}

\begin{abstract}
We study the effects of the $\Delta$(1232) resonance as an effective degree of freedom for charged and neutral pion photo-production on nucleons. Different observables have been calculated for these processes by using relativistic chiral perturbation theory up to $\mathcal{O}(p^3)$ in the $ \delta$ counting, thus, including pion loops. We compare our model with a large database containing the  available experimental data and  constrain some unknown low energy constants.
\end{abstract}

\bodymatter

\section{Introduction}
The pion photo-production processes on nucleons can be well described at low energies by using only nucleons and pions as the effective degrees of freedom. Baryon chiral perturbation theory (BChPT) reproduces fairly well these processes already at order $\mathcal{O}(p^4)$ both in heavy baryon and covariant approaches~\cite{Hilt:2013uf}. Despite the relative success of these calculations there are still some disagreements with data. Here, it will be shown that the effects of the $\Delta(1232)$ resonance mechanisms play a fundamental role and are essential to  correctly reproduce the energy dependence of various observables~\cite{Blin:2014rpa,Blin:2016itn}.

In this work,  we consider nucleons, pions and the lowest lying resonance, $\Delta$(1232), as the relevant degrees of freedom.  The calculation is carried out up to  and including $\mathcal{O}(p^3)$ in the $\delta$ power counting rule \cite{Pascalutsa:2002pi}. Furthermore, we use the extended on mass shell scheme (EOMS)~\cite{Fuchs:2003qc}, a covariant approach to BChPT.

This framework has been quite successful in the analysis of several processes, such as Compton scattering~\cite{Lensky:2009uv,Blin:2015era}, pion nucleon scattering~\cite{Alarcon:2012kn,Yao:2016vbz,Siemens:2016hdi} and  weak pion production off nucleons~\cite{Yao:2018pzc,Yao:2019avf}. Here, we focus on the different contributions of the $\Delta$(1232) in the charged and neutral pion photo-production on nucleons at low energies and we compare the results of our model with the $\Delta$-less approach.

\section{Theoretical model} 

The terms of the chiral Lagrangian contributing up to $\mathcal{O}(p^3)$ to the amplitude for $\gamma N \rightarrow \pi N'$ in the isospin limit can be classified according to their chiral order and the particles involved as
\begin{eqnarray}
\mathcal{L} = \mathcal{L}_N^{(1)} + \mathcal{L}_N^{(2)} + \mathcal{L}_{\pi \pi}^{(2)} +  \mathcal{L}_N^{(3)} +  \mathcal{L}_{\pi \pi}^{(4)} + \left( \mathcal{L}_{\Delta \pi N}^{(1)}+\mathcal{L}_{\Delta \gamma N}^{(2)} \right),
\label{eq:Lag}
\end{eqnarray}
where $\mathcal{L}^{(k)}$ indicates Lagrangian terms of  $\mathcal{O}(p^k)$.

For detailed definitions of  $\mathcal{L}_{\pi \pi}^{(2,4)}$  see~\cite{Gasser:1983yg,Gasser:1987rb}.

On the other hand, for the nucleonic terms, $\mathcal{L}_N^{(1,2,3)}$, we follow the conventions given in~\cite{Fettes:2000gb}. In this work, the  relevant LECs are the chiral limit axial coupling, the nucleon mass and the pion decay constant, $\{g,m,F_0\}$ in $\mathcal{L}_N^{(1)}$, as well as   $\{ c_1, c_6, c_7 \}$ in  $\mathcal{L}_N^{(2)}$ and the set $\{ d_8,d_9,d_{16},d_{18},d_{20},d_{21},d_{22} \}$ in $\mathcal{L}_N^{(3)}$.
Finally, we have the  $\Delta(1232)$ pieces, $\mathcal{L}_{\Delta \pi N}^{(1)}(h_A)$ and $\mathcal{L}_{\Delta \gamma N}^{(1)}(g_M)$, where $\{ h_A, g_M \}$ are respectively related to the strong and electromagnetic decay constants, $\Gamma_\Delta^{EM}$ and $\Gamma_\Delta^{strong}$, of the $\Delta(1232)$. See refs.~\cite{Pascalutsa:2007yg, Pascalutsa:2006up} for details. The  diagrams up to $\mathcal{O}(p^3)$ that contribute to the $\gamma N \rightarrow \pi N'$ reaction are the direct and crossed terms with a $\Delta$ as an intermediate state and leading to an $\mathcal{O}(p^{(5/2)})$ amplitude in the $\delta$ counting, as explained in~\cite{Navarro:2019iqj}.

For the pion photoproduction process 
the Lagrangian pieces shown in eq.~(\ref{eq:Lag})  generate different contributions at tree and one-loop level. The amputated amplitude order by order reads

\begin{equation}
    \begin{split}
        \hat{\mathcal{M}}& = \mathcal{M}^{(1)}_{\mbox{\footnotesize tree}} (\widetilde{g},\widetilde{m},F_0,M) + \mathcal{M}^{(2)}_{\mbox{\footnotesize tree}}(\widetilde{c}_1,\widetilde{c}_6,\widetilde{c}_7) + \mathcal{M}^{(5/2)}_{\mbox{\footnotesize tree}}(h_A,g_M) \\ 
        &+ \mathcal{M}^{(3)}_{\mbox{\footnotesize tree}}(l_3,l_4,d_8,d_9,d_{16},d_{18},d_{20},d_{21},d_{22}) + \widetilde{\mathcal{M}}^{(3)}_{\mbox{\footnotesize loop}},
    \end{split}
\label{eq:Meoms}    
\end{equation}
with $M$ the chiral limit pion mass from $\mathcal{L}_{\pi \pi}^{(2)}$, and $\widetilde{\mathcal{M}}^{(3)}_{\mbox{\footnotesize loop}}= \mathcal{M}^{(3)}_{\mbox{\footnotesize loop}} - \mbox{\small PCBT}$ the EOMS regularized loop amplitude. {\small PCBT} are the power counting breaking terms which are substracted through a redefinition of LECs at the lower order amplitudes, i.e., $\{ \widetilde{g}, \widetilde{m}, \widetilde{c}_1, \widetilde{c}_6,\widetilde{c}_7\}$ are the EOMS shifted LECs used to regularize the loop diagrams~\cite{Navarro:2019iqj}.

The loops appearing in the external legs of the Feynman diagrams are introduced by the Lehman-Symanzik-Zimmerman reduction formula~\cite{Lehmann:1954rq}. In particular, we use the EOMS regularized wave function renormalizations $\mathcal{Z}_N=1+\widetilde{\delta}_N^{(2)}$ and $\mathcal{Z}_\pi=1-2 l_4 m_\pi^2 +\widetilde{\delta}_{Z_\pi}^{(2)}$ as in \cite{Yao:2018pzc}. The UV divergences from the loop calculations are subtracted following the $\widetilde{MS}=\overline{MS}-1$ renormalization scheme. After the full renormalization procedure we implement further simplifications by expanding the chiral limit quantities in terms of its physical partners. Specifically, we take $\left\{ \widetilde{g}(g_A,d_{16}),\widetilde{m}(m_N,\widetilde{c}_1),F_0(F_\pi,l_4), M(m_\pi,l_3,l_4)\right\}$. In this way, we are able to write the renormalized amplitude $\mathcal{M}=\sqrt{\mathcal{Z}_\pi} \mathcal{Z}_N \hat{\mathcal{M}}$
as
\begin{equation}
    \begin{split}
         &\mathcal{M} =  \mathcal{M}^{(1)}_{\mbox{\footnotesize tree}} (g_A,m_N,F_\pi,m_\pi) + \mathcal{M}^{(2)}_{\mbox{\footnotesize tree}}(\widetilde{c}_6,\widetilde{c}_7) + \mathcal{M}^{(5/2)}_{\mbox{\footnotesize tree}}(h_A,g_M) \\ 
        &+ \mathcal{M}^{(3)}_{\mbox{\footnotesize tree}}(d_8,d_9,d_{18},d_{20},2d_{21}-d_{22}) + \widetilde{\mathcal{M}}^{(3)}_{\mbox{\footnotesize loops}} + (\widetilde{\delta}_{Z_N}^{(2)} + \frac{1}{2} \widetilde{\delta}_{Z_\pi}^{(2)})\mathcal{M}^{(1)}_{\mbox{\footnotesize tree}}.
    \end{split}
\label{eq:fullM}    
\end{equation}
Using this expression for the amplitude, we can compare with the  experimental data currently available~\cite{Navarro:2019iqj}, for various observables and constrain some of the unknown third order LECs~\cite{Navarro:2019iqj}.
\section{Results and discussion}
Here, we present the comparison among  two different models at $\mathcal{O}(p^3)$, one with the explicit inclusion of the $\Delta$ resonance and the other one without it.
For the experimental database,  we take into account a range of photon energies $E_\gamma^{lab} =155$ MeV - 215 MeV, including different observables such as the angular and total cross section, as well as the beam polarizabilities~\cite{Navarro:2019iqj}. In order to test the different models we minimize the $\chi$ squared taking the values for the  unknown LECs as the  fitting parameters. 
\\
In fact, many of the LECs are known from the analysis of other processes, in particular at $\mathcal{O}(p^3)$ we can set the values $\widetilde{c}_6 = 5.07(15)$~\cite{Bauer:2012pv,Yao:2019avf}, $\widetilde{c}_7=-2.68(08)$~\cite{Bauer:2012pv,Yao:2019avf}, $d_{18}=[ -0.20(80)$ GeV$^{-2}$]~\cite{Alarcon:2012kn}, $d_{22}=[5.20(02)$GeV$^{-2}$]~\cite{Yao:2017fym} and $h_A=2.87(03)$~\cite{Bernard:2012hb}. Consequently, the only fitting parameters are those given in Table~\ref{tab:fitting} for both models.\\
\begin{table}
\begin{minipage}[HT]{0.50\textwidth}

    {\includegraphics[width=2.2in]{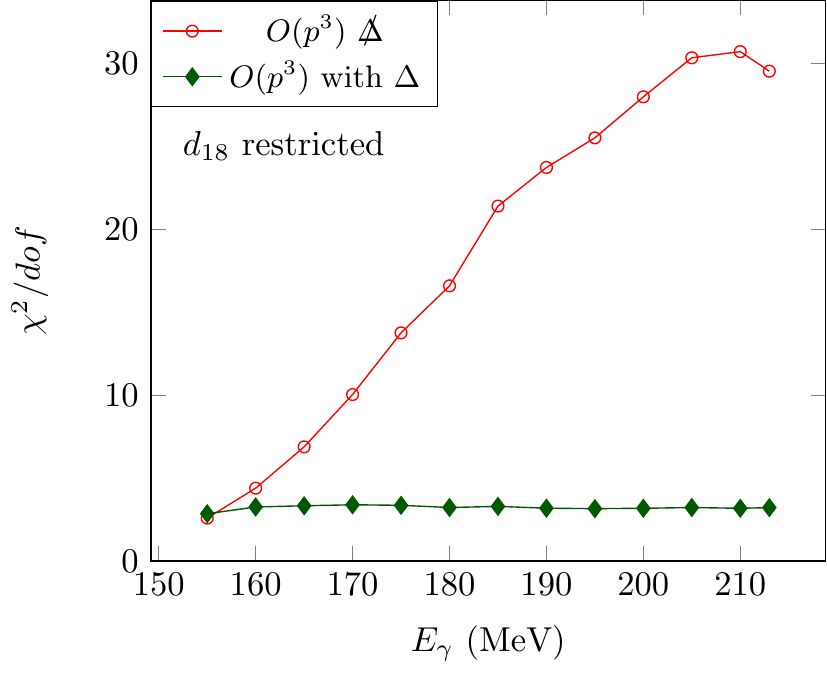}
    {\small Fig. 1. $\chi^2$ per degree of freedom as a function of the maximum photon energy of the data for the different studied models.}
    \label{fig:chi_beh}}

\end{minipage}\hspace{1.8cm}
\begin{minipage}[HT]{0.20\textwidth}
\tbl{The values of the LECs are dimensionless for $g_M$
and in units of GeV$^{-2}$ for d’s. Fit refers to the standard setting,
fit-$\slashed \Delta$ removes $\Delta$ mechanisms. Here $d_{18}$ is restricted to $-0.20(80)$, and therefore shown in boldface, while $g_M$ is left free}
{\begin{tabular}{@{}ccc@{}}\toprule
LEC & Fit  & Fit - $\slashed \Delta$ \\
\colrule
$d_8 + d_9$ & $1.16(01)$ & $3.53(01)$  \\
$d_8 - d_9$ & $1.09(18)$ & $5.31(24)$  \\
$d_{18}$ & $\mathbf{0.60}$ & $\mathbf{-1.00}$  \\
$d_{20}$ & $-0.74(17)$ & $-3.81(19)$ \\
$d_{21}$ & $4.32(14)$ & $6.98(15)$ \\
$g_M$ & $2.90(01)$ & ... \\
\colrule
$\chi^2_{TOT}/{d.o.f.}$& $\mathbf{3.22}$ & $\mathbf{29.5}$\\
\colrule
$\chi^2_{\pi 0}/{d.o.f.} $  & $3.58$  & $37.2$\\
$\chi^2_{\pi +}/{d.o.f.}$ & $1.89$ & $1.79$ \\
$\chi^2_{\pi -}/{d.o.f.}$ & $1.99$ & $1.90$ \\
\botrule
\end{tabular}
}
\label{tab:fitting}
\end{minipage}
\end{table}
According to the results in Table~\ref{tab:fitting}, the fit corresponding to the inclusion of $\Delta$ mechanisms results in a much better agreement with data than for the $\Delta$-less case. This result is mostly driven by the $\pi^0$ photoproduction channel, where $\chi^2$ is drastically reduced by the inclusion of the $\Delta$ at $\mathcal{O}(p^{5/2})$. On the other hand for $\pi^{\pm}$ channels the models show a similar agreement with or without the $\Delta$ inclusion. This is due to the fact that these channels are more sensitive to the lower orders. For that reason, considering higher order corrections does not  improve much the agreement. In Fig. 1, we show the quality of the fits as a function of the energy. It can be clearly seen the superior convergence of the model with the $\Delta$ resonance at the studied energies.\\
{\small
This research was supported by MINECO (Spain) and the ERDF (European Commission) grant No.~FIS2017-84038-C2-2-P, and SEV-2014-0398.}
\bibliography{weak}
\bibliographystyle{ws-procs9x6} 

\end{document}